# A note on Quarks and numbers theory

M. Hage-Hassan
Université Libanaise, Faculté des Sciences Section (1)
Hadath-Beyrouth

## Abstract

We express the basis vectors of Cartan fundamental representations of unitary groups by binary numbers. We determine the expression of Gel'fand basis of SU (3) based on the usual subatomic quarks notations and we represent it by binary numbers. By analogy with the mesons and quarks we find a new property of prime numbers.

## 1. Introduction

The applications of the SU (3) group theory in nuclear physics have occurred in particles and for description of nuclear collective properties and for the classification of elementary particles[1-2]. Representation of SU (3) and its generating function has been determined for a longtime[1,2]. Cartan found the fundamental representations of SU (n) and Gel'fand- Zeitlin [3] have found an orthogonal basis but the explicit determination of the expression of the basis of SU (n) are, very important in physics, not completely found. We want to study this problem using the generating function method [4].

In this work we observe that the vectors of the fundamental representations basis of SU (n) can be represented by binary numbers, the " binary basis ", very interesting for the determination of the generating function of SU(n) . We review the determination of the generating function of SU (3) by two different methods and we find the basis of Gel'fand based in terms of the well known Quarks indices [4-6]: {p, q}, I the isospin and its projection on the z axis, $I_z$, and the hypercharge Y. And we give also the correspondence between the binary basis, Gel'fand basis and the Quarks representations.

By analogy with the mesons and quarks we consider the primes{a}and it's supplements the primes { ā } with $n/2 \leq a < n$ and $a + \bar{a} = n$. We find n=5,7,10,16, 36 and 210.

The properties of SU (2) and the Gel'fand basis are treated in part two. We present in part three the fundamental representation and the binary basis. In the fourth, fifth and six parts we determine the generating function of SU (3) and the representation of quarks using the binary basis and the Gel'fand basis. In part seven we treat the analogy between The Baryons and Quarks and the prime numbers.

## 2. The basis of the unitary group

### *2.1 The SU(2) group*

**A- The SU(2) basis**

It's well known from Schwinger work [7] "on angular momentum" that the generators of angular momentum or SU(2) may be written in terms of creation and destruction operators of the two dimensional harmonic oscillator as:

$$J_+ = J_1 + iJ_2 = a_1^+ a_2,\ J_- = J_1 - iJ_2 = a_2^+ a_1,\ J_3 = [a_1^+ a_1 - a_2^+ a_2]/2 \quad (2,1)$$

With $[a^+, a] = 1, and\ \ [J_3, J_\pm] = J_\pm, [J_+, J_-] = 2J_3$

The basis is:

$$|\, jm\rangle = \frac{a_1^{+(j+m)} a_2^{+(j-m)}}{\sqrt{(j+m)!(j-m)!}} |0,0\rangle = \phi_{jm}(a^+)|0,0\rangle \quad (2,2)$$

With $j \geq |m|,\ j = 0,\ 1/2,\ 1,\ 3/2,\ 2,\ldots$

$$J_\pm |\, jm\rangle = \sqrt{j(j+1) - m(m \pm 1)}\ |\, jm \pm 1\rangle$$

$J_+$ and $J_-$ are the raising and lowering operators of SU(2).

And $$|\, jm\rangle = \sqrt{\frac{(j \pm m)!}{(2l)!(l \mp m)!}}\ J_\mp^{j \mp m} |\, j \pm j\rangle, \quad (2,3)$$

$|\, j \pm j\rangle$ are the maximal et minimal states of SU(2).

**B- The generating function of SU(2)**

The generating function of SU(2) is give by:

$$|G(z,q)\rangle = \sum_{jm} \varphi_{jm}(z) |\, jm\rangle = \exp[z_1 a_1^+ + z_2 a_2^+]|0,0\rangle, \quad (2,4)$$

$$\varphi_{jm}(z) = \frac{z_1^{j+m} z_2^{j-m}}{\sqrt{(j+m)!(j-m)!}}$$

$\varphi_{jm}(z)$ is the basis of the analytic Hilbert space, the Fock or the Fock-Bargmann(F-B) space, with the Gaussian measure:

$$d\mu(z) = \exp(-z \cdot \bar{z}) dz_1 dz_2,\ dz_i = dx_i dy_i,\ z_i = x_i + iy_i$$

It's important to note that $z_1$ and $z_2$ have the same powers of $J_+$ and $J_-$ (2,3).

### *2.2 The Gel'fand basis of the unitary groups*

Using the "Weyl's branching law" Gelfand-Zeitlin [8] introduced the basis of representation of U (n), function of $n(n+1)/2$ integers numbers.

The Gel'fand basis $|(h)_n\rangle$ is:

$$|(h)_n\rangle = \left| \begin{matrix} h_{1n} & & h_{2n} & & \cdots & & h_{nn} \\ & h_{1n-1} & & & \cdots\cdots & & \\ & & h_{12} & & h_{22} & & \\ & & & h_{11} & & & \end{matrix} \right\rangle = \left| \begin{matrix} [h]_n \\ (h)_{n-1} \end{matrix} \right\rangle = \left| \begin{matrix} [h]_n \\ [h]_{n-1} \\ (h)_{n-2} \end{matrix} \right\rangle \cdots \quad (2,5)$$

With $h_{i,j} \geq h_{i,j-1} \geq h_{i+1,j},\ j = 2,\ldots,n\ \ et\ \ i = 1,\ldots, j.$

And $h_{nn}=0$ for SU(n)

### *2.3 The Weyl dimension formula*

The dimension of subspaces $[h_{\mu\nu}]$ is finite and given by the Weyl formula:

$$d_{[h_{\mu\nu}]} = \left[ \prod_{i<j} (p_{in} - p_{jn}\} \right] / [1!2!\cdots(n-1)!] \quad (2,6)$$

With $p_{in} = h_{in} + n - i$

### *2.4 The maximal and minimal states*

We associate to each state $|(h)_n\rangle$ a vector or weight vector which has components:

$$\omega(h) = (\omega_{1n}(h),\ \omega_{2n}(h), ...\ \omega_{nn}(h)).$$

With
$$\omega_{in} = (\sum_{j=1}^{i} h_{j,i} - \sum_{j=1}^{i-1} h_{j,i-1})$$

A weight $\omega(h')$ is higher than a weight $\omega(h)$ if the first nonzero component in the difference $\omega(h') - \omega(h)$ is positive.

We deduce simply that there are a maximal and a minimal vector as SU(2).

### *2.5 Explicit expression of Gel'fand basis vectors*

By analogy with SU(2), Nagel and Moshinsky [9] proved that the basis $|(h)_n\rangle$ may be deducted from the minimal or the maximal vectors by applying the raising operators $R_\lambda^\mu$ or the lowering operators $L_\lambda^\mu$ and find the explicit expressions of these operators.

We write:

$$|(h)_n\rangle = N \prod_{\lambda=2}^{n}\prod_{\mu=1}^{k-1} (L_\lambda^\mu)^{L_\lambda^\mu} \left| \begin{matrix} [h]_n \\ (\max)_{n-1} \end{matrix} \right\rangle = N' \prod_{\lambda=2}^{n}\prod_{\mu=1}^{k-1} (R_\lambda^\mu)^{R_\lambda^\mu} \left| \begin{matrix} [h]_n \\ (\min)_{n-1} \end{matrix} \right\rangle$$

With
$$L_\lambda^\mu = h_{\mu,\lambda} - h_{\mu,\lambda-1} \quad , \ R_\lambda^\mu = h_{\mu,\lambda-1} - h_{\mu+1,\lambda} \qquad (2,7)$$

N and N 'are the normalization constants.

## 3. The fundamental representations and the Binary numbers

### 3.*1 The fundamental representations:*

E. Cartan proved that any arbitrary irreducible representation of U(n) can be expressed in terms of a set of subspaces called the fundamental representations [6].

The fundamental representations of U (n) are the irreducible subspaces $[h]_n$:

$$[1,0,\cdots,0]_n,\ [1,1,\cdots,0]_n,\ \cdots,[1,1,\cdots,1]_n \qquad (3,1)$$

The dimension of the subspace:

$$[\overbrace{1,1,1,..,1}^{p},0,...,0]_n = F_{(n,p)}$$

is $C_n^p$ then dim $F_{(n,n)} = 1$.

It is simple to prove that that the total number of basis vectors of the fundamental representations of U(n) is $2^n$-1 and $2^n$-2 for SU(n).

### *3.2 The binary numbers of the fundamental representations:*

We observe that the vectors basis of the fundamental representations may be expressed in terms of the binary numbers so we called it the binary fundamentals basis. And it is easy to establish the correspondence between these binary basis and the fundamentals representation of Gel'fand basis.

We write as an example:

| | | | | | | | | | | | | | | | | | | | |
|---|---|---|---|---|---|---|---|---|---|---|---|---|---|---|---|---|---|---|---|
| $F_{(2,1)}$ | 1 | 0 | & | 0 | 1 | | | | | | | | | | | | | | |
| $F_{(3,1)}$ | 1 | 0 | 0 | & | 0 | 1 | 0 | & | 0 | 0 | 1 | | | | | | | | |
| $F_{(3,2)}$ | 0 | 1 | 1 | & | 1 | 0 | 1 | & | 1 | 1 | 0 | | | | | | | | |
| $F_{(4,1)}$ | 1 | 0 | 0 | 0 | & | 0 | 1 | 0 | 0 | & | 0 | 0 | 1 | 0 | & | 0 | 0 | 0 | 1 |
| $F_{(4,2)}$ | 1 | 1 | 0 | 0 | & | 1 | 0 | 1 | 0 | & | 1 | 0 | 0 | 1 | | | | | |
| $F_{(4,2\}}$ | 0 | 0 | 1 | 1 | & | 0 | 1 | 0 | 1 | & | 0 | 1 | 1 | 0 | | | | | |
| $F_{(4,3)}$ | 0 | 1 | 1 | 1 | & | 1 | 0 | 1 | 1 | & | 1 | 1 | 0 | 1 | & | 1 | 1 | 1 | 0 |

***Table 1***

### *3.3 The representations of the binary basis in the analytic Hilbert space*

Let $(z_i^j),\ i,j=1,...,n$ a matrix of complexes numbers and we consider the minors of this matrix: We associate to each miner $\Delta_{i_1..i_l}^{12..l}$ of the matrix $(z_i^j)$ i, j=1,.., n a table of n-boxes numbered from 1 to n. We put "one" in the boxes $i_1, i_2, \ldots, i_l$ and zeros elsewhere.

$$\begin{matrix} & & 1 & 2 & \ldots & i_1 & \ldots & i_k & \ldots & n \\ \Delta_{[i]}^k = \Delta_{i_1\ldots i_k}^{12\ldots k} & = & |0 & 0 & \ldots & 1 & \ldots & 1 & \ldots & 0| \end{matrix}$$

We denote these orthogonal basis of F-B space by:

$$\Delta_{[i]}^k(z) = \left\{\Delta_n^i = \Delta_{i_1 i \ldots i_l}^{12\ldots k}(z) = F_{(n,i)}(z)\right\}$$

For SU(3) we write:

$$\Delta_3^1 = z_1^1, \Delta_3^2 = z_2^1, \Delta_3^3 = z_3^1,$$

$$\Delta_3^4 = \begin{vmatrix} z_2^1 & z_3^1 \\ z_2^2 & z_3^2 \end{vmatrix}, \Delta_3^5 = \begin{vmatrix} z_1^1 & z_3^1 \\ z_1^2 & z_3^2 \end{vmatrix}, \Delta_3^6 = \begin{vmatrix} z_1^1 & z_2^1 \\ z_1^2 & z_2^2 \end{vmatrix}, \qquad (3,2)$$

The Gel'fand representation can be written in the analytic Hilbert or F-B space by:

$$\Gamma_n(\Delta(z)) = \langle (h)_n \| \Delta(z) \rangle$$

# 4. Schwinger approach and generating function of SU(3)

### *4.1 The coupling in the angular momentum*

We emphasize only the Schwinger couplings formula [5,7]:

$$G(\alpha, z) = \exp[\{\alpha_3[x,y] + \alpha_1(z,y) + \alpha_2(x,z)\}] =$$

$$\sum_{j1j2jm} \frac{(\alpha_3)^{((j1+j2)-j)}}{\sqrt{((j_1+j_2)-j)!}} \varphi_{j(j1-j2)}(\alpha)\varphi_{(j1j2)jm}(x,y) =$$

$$\exp[\{\alpha_3[\frac{\partial}{u},\frac{\partial}{v}] + \alpha_2[z_1\frac{\partial}{u_1} + z_2\frac{\partial}{u_2}] + \alpha_1[z_1\frac{\partial}{v_1} + z_2\frac{\partial}{v_2}]]\}\exp[(u,y)+(v,x)] \qquad (4,1)$$

$\varphi_{(j1j2)jm}(x,y)$ is the coupled function of $\varphi_{j_1 m_1}(x)$ and $\varphi_{j_2 m_2}(z)$.

This formula determine simply the coupling of several angular momentum and the representation of SU (3) and for simplicity we use the F-B space.

### *4.2 The basis of the group SU (2) $\subset$ SU (3)*

Let $D_{[\lambda,\mu]}$ the space of homogeneous polynomials and $V^{\lambda\mu}_{(t,tz,y)}(z^1,z^2)$ is the orthogonal basis with:

$$z^{(1)} = (z_1^1, z_2^1, z_3^1),\ z^{(2)} = (z_1^2, z_2^2, z_3^2),\ z_i^j \in C \qquad (4,2)$$

And $\qquad T_{ij} = \sum_k z_k^i(\partial / \partial z_k^i)$

The space is homogeneous then

$$T_{11}V^{\lambda\mu}_{(\alpha)} = (\lambda+\mu)V^{\lambda\mu}_{(\alpha)}, \qquad T_{22}V^{\lambda\mu}_{(\alpha)} = (\mu)V^{\lambda\mu}_{(\alpha)} \qquad (4,3)$$

The vectors $V^{\lambda\mu}_{(t,tz,y)}(z^1,z^2)$ are eigenfunctions of the Casimir operator of the second order $\vec{T}^2$, the projection of $\vec{T}$ on the z axis and the hypercharge Y. The eigenvalues of these operators are respectively t (t + 1), $t_z$ and the triple of the hypercharge quantum number y. The numbers $t, t_z$ are the isospin and the component of isospin on the z axis.

We have:

$$YV^{\lambda\mu}_{(\alpha)} = yV^{\lambda\mu}_{(\alpha)}, \qquad T_z V^{\lambda\mu}_{(\alpha)} = t_z V^{\lambda\mu}_{(\alpha)}$$

And

$$\vec{T}^2 V^{\lambda\mu}_{(\alpha)} = t(t+1)V^{\lambda\mu}_{(\alpha)}. \qquad (4,4)$$

the condition of Young tableau on $V^{\lambda\mu}_{(\alpha)}$ imposes the further condition:

$$T_{12}V^{\lambda\mu}_{(\alpha)}(z^1,z^2) = 0 \qquad (4,5)$$

The expression of $V^{\lambda\mu}_{(\alpha)}(z^1,z^2)$ is well known [1,5] and we give only the result:

$$V^{\lambda\mu}_{(t,tz,y)}(z^1,z^2) = N1(\lambda\mu;\alpha)(-1)^s \times \sum_k \binom{m}{k} \frac{(\mu-s)!r!}{(\mu-s-k)![r-(n-k)]!} \qquad (4,6)$$
$$\times (\Delta_3^1)^{r-(m-k)}(\Delta_3^2)^{m-k}(\Delta_3^3)^{\lambda-r}(\Delta_3^4)^k(-\Delta_3^5)^{\mu-s-k}(\Delta_3^6)^s$$

$$N1(\lambda\mu;\alpha) = \left\{ \frac{(\lambda+1)!(\mu+r-s+1)!}{r!s!(\mu-s)!(\lambda-r)!(\mu+r+1)!(\lambda+\mu-s+1)!} \times \frac{(2t-m)!}{(2t)!m!} \right\}^{\frac{1}{2}}$$

And $\qquad y = -(2\lambda+\mu)+3(r+s),\ 0 \le r \le \lambda, \qquad (4,7)$

$$t = \frac{\mu}{2} + \frac{r-s}{2},\quad 0 \le s \le \mu,\quad \text{And} \quad t_z = t-m,\quad m = 0,1,\ldots,2t.$$

### *4.3 The generating function of SU(3)*

The vectors $V^{\lambda\mu}_{(t,tz,y)}(z^1,z^2)$ are the elements of the product spaces $D_{t_1} \otimes D_{t_2} \otimes D_{t_3}$ which has the basis vectors:

$$\varphi_{j_1m_1}(z_1^1,z_2^1)\varphi_{j_2m_2}(z_1^2,z_2^2)\varphi_{j_3m_3}(z_3^1,z_3^2). \qquad (4,8)$$

But

$$T_+ = \Im_+ + z_3^1\frac{\partial}{\partial z_3^2}, \qquad T_- = \Im_- + z_3^2\frac{\partial}{\partial z_3^1}, \qquad T_z = \Im_3 + \frac{1}{2}(z_3^1\frac{\partial}{\partial z_3^1} - z_3^2\frac{\partial}{\partial z_3^2})$$

it is easy to verify that the generators $\Im_+, \Im_-$ *et* $\Im_3$ are the generators of SU (2).

We following Schwinger coupling method for the determination of $\{V^{\lambda\mu}_{(\alpha)}\}$ by:

a-The first coupling: $\varphi_{j_1 m_1}(z^1_1, z^1_2)\varphi_{j_2 m_2}(z^2_1, z^2_2)$

b-We apply the result obtained the second coupling

$$\varphi_{j(j1-j2)}(\alpha)\varphi_{j_3 m_3}(z^1_3, z^2_3),$$

c- We apply then the Young pattern condition we obtain the generating function of SU(3) [1-5]. But we write it in a simple form by:

$$G((x,y,u),z) = \exp[\vec{f}.\vec{\Delta}^1 + \vec{g}.\vec{z}^{(12)}] = \sum_{\lambda\mu\, tt_0 y} \varphi^{(\lambda\mu)}_{(t,t_z,y)}(\vec{f},\vec{g})\, V^{(\lambda\mu)}_{(t,t_z,y)}(z^{(1)}, z^{(12)}) \quad (4,9)$$

With $\vec{f} = (x_1\xi, x_1\eta, x_2), \ \vec{g} = (y_1\eta, -y_1\xi, y_2)$

And $\vec{z}^{(i)} = (z^i_1, z^i_2, z^i_3), \ \vec{z}^{(ij)} = \vec{z}^{(i)} \times \vec{z}^{(j)}$

We have

$$\varphi^{(\lambda\mu)}_{(t,t_z,y)}(\vec{f},\vec{g}) = N[(\lambda\mu),(\alpha)](x_1^r x_2^{\lambda-r} y_1^{(\mu-s)} y_2^s)(\xi^{(t+t_z)}\eta^{(t-t_z)}) \quad (4,10)$$

And $$N[(\lambda\mu),(\alpha)] = (-1)^s \sqrt{\frac{(\mu+r+1)!(\mu+\lambda-s+1)!}{(\lambda+1)!(2t+1)\lambda!\,[r!(\lambda-r)!s!(\mu-s)!(t+t_z)!(t-t_z)!}}$$

## 5. Generating function of SU(3) and Gel'fand basis

### *5.1 Generating function of harmonic oscillator and the Gel'fand basis*

The generating function of the oscillator in terms of Gel'fand indices is:

$$\left|G(z,q)\right\rangle = \sum_{h_{11}} \frac{(y^1_1)^{h_{11}}}{\sqrt{(h_{11})!}}\left|(h_{11})\right\rangle = \exp[y^1_1 a^+_1]\left|0\right\rangle \quad (5,1)$$

### *5.2 Generating function of SU(2) and the Gel'fand basis*

We express the generating function of SU(2) in terms of Gel'fand indices by:

$$\left|G(z,q)\right\rangle = \sum_{h_{11},h_{12}} \frac{(x^1_2)^{h_{12}-h_{11}}(y^1_2)^{h_{11}}}{\sqrt{(h_{12}-h_{11})!(h_{11})!}}\left|\begin{smallmatrix} h_{12} & & 0 \\ & h_{11} & \end{smallmatrix}\right\rangle = \exp[y^1_2 a^+_1 + x^1_2 a^+_2]\left|0,0\right\rangle \quad (5,2)$$

With $2j = h_{12}, \ j+m = h_{11}, \ h_{22} = 0$.

We have already noted that the powers of $x^1_2$ and $y^1_2$ have the same powers of raising and lowering operators (2,5).

### *5.3 The generating function of SU (3) and the Gel'fand basis*

Generalizing the generating functions of the oscillator and SU (2) to SU (3) we write:

$$\sum_{\lambda\mu} \prod_{\lambda=2}^{n}\prod_{\mu=1}^{\lambda-1} A_{\lambda\mu}\left((x^\mu_\lambda)^{h_{\mu\,\lambda}-h_{\mu\,\lambda-1}}(y^\mu_\lambda)^{h_{\mu\,\lambda-1}-h_{\mu+1\,\lambda}}\right)\Gamma_n(\Delta(z))$$

$$= \exp[\sum_i \varphi^n_i(x,y) F_{(n,i)}(z)], \ n = 3 \quad (5,3)$$

We note that the powers of the parameters x, y are the same powers of raising and lowering operators of SU(n) (2,6).
Comparing the two generating functions (4,10) et (5,3) of SU (3) we find:

$$p=\lambda+\mu,\ q=\mu,\ I=t \text{ et } I_z=t_z$$

$$t-t_z= h_{12}-h_{11},\quad t+t_z= h_{11}-h_{22},$$

And $\quad \lambda-r=h_{13}-h_{12},\quad r= h_{12}-h_{23},\quad \mu -s= h_{23}-h_{22},\quad s= h_{22}.$ (5,4)

After solving the system we write the $(h)_3$ of Gel'fand basis by:

$$\begin{pmatrix} \lambda+\mu & & \mu & & 0 \\ & I+\frac{Y}{2}+\frac{(\lambda+2\mu)}{3} & & -I+\frac{Y}{2}+\frac{(\lambda+2\mu)}{3} & \\ & & I_z+\frac{Y}{2}+\frac{(\lambda+2\mu)}{3} & & \end{pmatrix} \qquad (5,5)$$

Thus we find the same result obtained by Baird and Biedenharn by another method [5]. So we find the Gel'fand basis in the Quarks notations.
The generalization of the generating function for SU (n) is already studied [3-4].

## 6. Quarks and Binary numbers

The fundamental representations of quarks and antiquarks [2,10] are represented by:

a- The quarks states are [3] : $|I_z,\ Y>$
with quantum numbers: $(I_z, Y) =$ (1/2,1/3), (-1/2, 1/3), (0, -2/3)

b- The antiquarks states are$[\overline{3}]$: $|\overline{I_z},\ \overline{Y}>$
with quantum numbers: $(I, Y) =$ (-1/2, -1/3), (1/2, -1/3), (0, 2/3)

| SU(2) \| i = | 1 | 2 |
|---|---|---|
| **Binary basis $F_{(2,i)}$** | 1 \| 0 | 0 \| 1 |
| **Gel'fand basis** | $\begin{pmatrix}1 & & 0\\ & 1 & \end{pmatrix}$ | $\begin{pmatrix}1 & & 0\\ & 0 & \end{pmatrix}$ |
| $\phi_i^2$ | $y_2^1$ | $x_2^1$ |
| **isospin $I_z$** | 1/2 | -1/2 |

| SU(3) \| i = | 1 | 2 | 3 | 4 | 5 | 6 |
|---|---|---|---|---|---|---|
| **Binary basis $F_{(3,i)}$** | 1 \| 0 \| 0 | 0 \| 1 \| 0 | 0 \| 0 \| 1 | 0 \| 1 \| 1 | 1 \| 0 \| 1 | 1 \| 1 \| 0 |
| **Gel'fand basis** | $\begin{pmatrix}1 & & 0 & & 0\\ & 1 & & 0 & \\ & & 1 & & \end{pmatrix}$ | $\begin{pmatrix}1 & & 0 & & 0\\ & 1 & & 0 & \\ & & 0 & & \end{pmatrix}$ | $\begin{pmatrix}1 & & 0 & & 0\\ & 0 & & 0 & \\ & & 0 & & \end{pmatrix}$ | $\begin{pmatrix}1 & & 1 & & 0\\ & 1 & & 0 & \\ & & 0 & & \end{pmatrix}$ | $\begin{pmatrix}1 & & 1 & & 0\\ & 1 & & 0 & \\ & & 1 & & \end{pmatrix}$ | $\begin{pmatrix}1 & & 1 & & 0\\ & 1 & & 1 & \\ & & 1 & & \end{pmatrix}$ |
| $\phi_i^3$ | $y_2^1 y_3^1$ | $x_2^1 y_3^1$ | $x_3^1$ | $x_2^1 x_3^2$ | $y_2^1 x_3^2$ | $y_3^2$ |
| **Quarks ($I_z$, Y)** | (1/2,1/3) | (-1/2,1/3) | (0,-2/3) | (-1/2,-1/3) | (1/2, -1/3) | (0, 2/3) |

***Table 2.*** binary basis, Gel'fand basis and Quarks notations of SU(3).

**Remarque: calculus of the coefficients $\phi_i^n(x,y)$**

The coefficients $\phi_i^n(x,y)$ may be written as product of parameters $y_\lambda^\mu = y(\lambda,\mu)$ and $x_\lambda^\mu = x(\lambda,\mu)$. We determine the indices of these parameters by using the following rules:

a- We associate to each "one" which appeared after the first zero a parameter $y(\lambda,\mu)$ whose index $\lambda$ are the number of boxes and $\mu$ the number of "one" before him, plus one.

b- We associate to each zero after the first "one" a parameter $x(\lambda,\mu)$ whose index $\lambda$ is the number of boxes and $\mu$ the number of "one" before him.

## 7. Quarks and the prime numbers

By analogy with the mesons, quarks and anti quarks we seek to find the numbers n with the conditions: for any prime a, n/2≤ a<n = 2,3,…etc. there are a prime supplements ā of a with a + ā = n.

If n = 12 the primes{a} are 11, 7 and their supplements{ā}: 1, 5 then twelve must be eliminate because one is not in the list of Almanac as prime.

| | | | | | | | | |
|---|---|---|---|---|---|---|---|---|
| 5 | | 3,2 | | | | | | |
| 7 | | 5,2 | | | | | | |
| 10 | | 5,5 | 3,7 | | | | | |
| 16 | | 5,11 | 3,13 | | | | | |
| 36 | | 17,19 | 13,23 | 7,29 | 5,31 | | | |
| 210 | | 103,107 | 101,109 | 97,113 | 83,127 | 79,131 | 73,137 | 71,139 |
| | | 61,149 | 59,151 | 53,157 | 47,136 | 43,167 | 37,173 | 31,179 |
| | | 29,181 | 19,191 | 17,193 | 13,197 | 11,199 | | |

Table 3: calculation of (ā, a) with n≤1000.